\documentstyle[twoside,fleqn,espcrc2,epsfig]{article}

\title{High temperature meson propagators with domain-wall quarks.\thanks
       {Talk presented by D.~K.~Sinclair. Supported by DOE contract
        W-31-109-ENG-38.}}

\author{J.-F.~Laga\"{e}$^{\rm a}$ and D.~K.~Sinclair\address{HEP Division, 
        Argonne National Laboratory, 9700 South Cass Avenue, Argonne, 
        IL 60439, USA}}

\begin{document}

\begin{abstract}
We study the chiral properties of domain-wall quarks at high temperatures on
an ensemble of quenched configurations. Low lying eigenmodes of the Dirac
operator are calculated and used to check the extent to which the Atiyah-Singer
index theorem is obeyed on lattices with finite $N_5$. We calculate the
connected and disconnected screening propagators for the lowest mass scalar
and pseudoscalar mesons in the sectors of different topological charge and note
that they behave as expected. Separating out the would-be zero eigenmodes
enables us to accurately estimate the disconnected propagators with far less
effort than would be needed otherwise.
\end{abstract}

\maketitle

\section{Introduction}

We use Shamir's formulation \cite{shamir} of lattice domain-wall fermions, which
have exact chiral flavour symmetry for infinite 5th dimension ($N_5$), to
study the chiral properties of quarks in high temperature QCD and their relation
to topology.

The chirally symmetric Dirac operator should have a zero mode associated with
each instanton, and obey the Atiyah-Singer index theorem. In addition, these
modes yield the disconnected contributions which distinguish the $\pi$ from
the $\eta'$ screening propagators and the $\sigma(f_0)$ from the $\delta(a_0)$
propagators, and give related contributions to the connected propagators.
Since we use quenched configurations, we cannot determine whether the
anomalous $U(1)$ axial symmetry is restored in the high temperature phase of
the massless $N_f=2$ theory, but we can check relations required for the
$U(1)$ axial symmetry to remain broken.

We measured eigenmodes of the domain-wall Dirac operator as a function of
$N_5$ on a set of quenched configurations in the high temperature phase. We
check the Atiyah-Singer theorem for $N_5=10$ and calculate connected and
disconnected scalar and pseudoscalar screening propagators. Early results were
presented at Lattice'98 \cite{lagae98}. Section 2 describes our analysis and
tests of the index theorem. The meson screening propagators are discussed in
section 3. Section 4 gives our discussions and conclusions.

\section{Domain-wall quarks at high temperatures}

For our study of domain-wall quarks at high temperatures we use $16^3 \times
8$ quenched configurations at $\beta=6.2$ (170), $\beta=6.1$ (100) and
$\beta=6.0$ (100). (At $N_t=8$, $\beta_c \approx 6.0$.) On each configuration,
we estimated the topological charge by the cooling method.

We studied eigenvalue trajectories of the hermitian Wilson Dirac
operator $\gamma_5 D_W$ as the bare mass was varied. These vanish at
would-be zero modes associated with instantons \cite{smitvink}. Based on these
studies we chose the Shamir domain-wall mass parameter $M=1.7$ for all 3
$\beta$ values.

We calculated the lowest 2 eigenmodes of the hermitian domain-wall Dirac
operator $\gamma_5 {\cal R} D_{dw}$ for all $\beta=6.2$ configurations 
with non-trivial topology and for all $\beta=6.1$ and $\beta=6.0$
configurations, for $N_5=4$, $6$, $8$ and $10$. (A third $N_5=10$ eigenmode was
calculated for each $\beta=6.0$ configuration with topological
charge $\pm 3$.)

For $\beta=6.2$, there is a clear separation between would-be zero modes whose
eigenvalues decrease exponentially with $N_5$ and non-zero modes which rapidly
approach a constant. By $\beta=6.0$ the situation is less clear. For
$\beta=6.2$, $N_5=10$ appears adequate.

\begin{figure}[htb] 
\epsfxsize=3in
\centerline{\epsffile{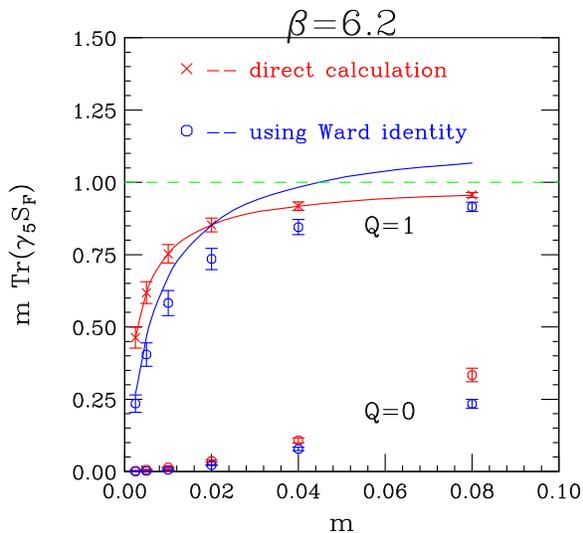}}
\caption{$m{\rm Tr}(\gamma_5 S_F)$ as a function of mass for $\beta=6.2$. The
         solid lines for $Q=1$ are the contributions from the lowest eigenmode.
                                      \label{fig:AS}}
\end{figure}

We calculate the chiral condensates $\langle\bar{q}q\rangle$ and
$\langle\bar{q}\gamma_5 q\rangle$ for the case $N_5=10$, using an eigenmode
enhanced stochastic estimator. For each configuration, exact chiral symmetry
requires $\langle\bar{q}\gamma_5 q\rangle$ to obey the Atiyah-Singer index
theorem:
\begin{equation}
m\sum_x\langle\bar{q}(x)\gamma_5 q(x)\rangle_U = Q_U.
\end{equation}
As figure~\ref{fig:AS} shows, the index theorem at $\beta=6.2$ is well 
approximated down to masses comparable to the eigenvalue of 
$\gamma_5 {\cal R} D_{dw}$ at $m=0$ of smallest magnitude.

\section{Meson screening propagators.}

Screening propagators, i.e. propagators for spatial separations, describe the
excitations of hadronic matter and the quark-gluon plasma at finite
temperature. We measure the connected and disconnected parts of these
propagators for scalar and pseudoscalar mesons with zero momenta transverse to
their separation ($Z$).

The connected parts of these propagators are measured with a noisy point source
on 1 $z$ slice for $Q=0$ and on each $z$ slice for $Q \neq 0$. The
disconnected parts of these propagators are measured separating out the
contributions of the eigenmodes (projected on to the domain walls) and
approximating the remainder of the required traces with a stochastic estimator.
For a discussion of stochastic estimators for staggered quarks see\cite{kilcup}.

\begin{figure}[htb] 
\epsfxsize=3in
\centerline{\epsffile{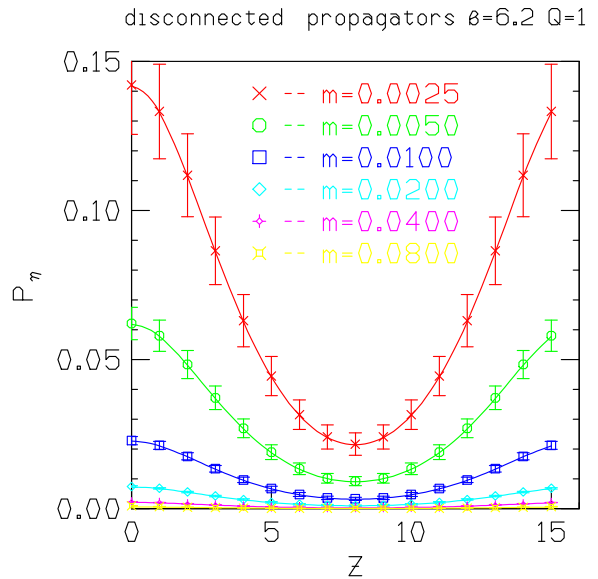}}
\caption{The disconnected contribution to the $\eta'$ propagator at $\beta=6.2$.
         The corresponding graph for the $\sigma$ is virtually identical.
                           \label{fig:disconnected}}
\end{figure}

The $\beta=6.2$ disconnected propagators behave as expected. For $Q=1$, the
contribution is sizeable and increases as quark mass decreases (it should
diverge as $1/m^2$ if chiral symmetry were exact), as shown in
figure~\ref{fig:disconnected}.
For $Q=0$ the contribution is extremely small compared with the connected
contribution (on the scale of figure~\ref{fig:disconnected} all points
would lie on the horizontal axis). This is consistent with the expectation
that the $Q=0$ contribution vanishes in the chiral limit.

For the connected propagators, we consider the combinations $\frac{1}{2}(P_\pi
\pm P_\delta)$ (Note: the connected part of the $\eta'$ propagator is the
$\pi$ propagator and the connected part of the $\sigma$ propagator is the
$\delta$ propagator). For $Q=0$, flavour chiral symmetry restoration and the
absence of a disconnected contribution would make $\frac{1}{2}(P_\pi +
P_\delta)$ vanish in the chiral limit. The observed value is very small. The
other combination, $\frac{1}{2}(P_\pi - P_\delta)$, 
is non-vanishing and has a finite chiral limit, as
expected.

\begin{figure}[htb]                                                            
\epsfxsize=3in                                                             
\centerline{\epsffile{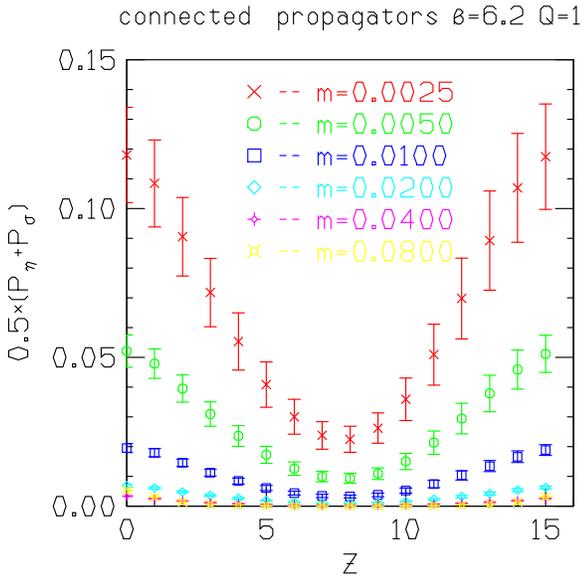}}                             
\caption{The connected propagator $\frac{1}{2}(P_\pi + P_\delta)$ for $Q=1$ at
         $\beta=6.2$.\label{fig:p+d}}                                         
\end{figure}

In the $Q=1$ sector, $\frac{1}{2}(P_\pi + P_\delta)$ should equal the 
disconnected part of the $\sigma$ and $\eta'$ propagators in the chiral limit,
for chiral flavour symmetry restoration with broken $U(1)$ axial symmetry to
be possible for $N_f=2$ \cite{kls}. $\frac{1}{2}(P_\pi + P_\delta)$ is plotted 
in figure~\ref{fig:p+d}.
Figures \ref{fig:p+d} and \ref{fig:disconnected} are consistent with this
expectation.

\section{Discussion and conclusions.}

For quenched lattice QCD at $N_t=8$, domain-wall quarks with $N_5 \geq 10$
are a good approximation to chiral quarks at $\beta=6.2$ and suggest an approach
to chiral symmetry which is exponential in $N_5$. The index theorem is well
approximated and the $\pi$, $\sigma$, $\eta'$ and $\delta$ propagators show the
correct behaviour for the restoration of chiral $SU(2) \times SU(2)$ flavour
symmetry for $N_f=2$. The disconnected parts of the $\sigma$ and $\eta$ 
propagators come entirely from the $Q \neq 0$ sector and have the behaviour
needed to give a finite contribution when the $N_f=2$ determinant is included.
Isolating the lowest eigenmodes greatly improves our stochastic estimation of
traces.

Our preliminary results for $\beta=6.0$ indicate a more complex behaviour. It
is not yet clear whether this indicates that $N_5=10$ is too small, or that
the condensation of instanton-antiinstanton pairs, as observed by Heller et al.
for $N_t=4$ using Ginsparg-Wilson fermions \cite{heller}, is playing an
important role.

A dynamical simulation is needed to determine if, for $N_f=2$, the $U(1)$ axial
symmetry remains broken above the transition. Such simulations are being
performed by the Columbia group at $N_t=4$ \cite{vranas}, and preliminary
indications are that it does remain broken. These pioneering results are
limited by the strong couplings dictated by $N_t=4$, and use of heavier quarks
than would be desirable. To push closer to the chiral limit probably requires
isolating the lowest eigenmode(s) in the simulations.

These calculations were performed on the C90 and J90's at NERSC.

\end{document}